\documentclass[prd,twocolumn,showpacs,superscriptaddress,nofootinbib]{revtex4}
\usepackage{graphicx}
\usepackage{dcolumn}
\usepackage{bm}
\usepackage{amssymb}
\usepackage{amsmath}
\usepackage{hhline}
\usepackage{color}

\begin{document}

\title{Reconciling the 2 TeV Excesses at the LHC in a Linear Seesaw Left-Right Model}

\author{Frank F. Deppisch}
\email{f.deppisch@ucl.ac.uk}
\affiliation{Department of Physics and Astronomy, University College London, London WC1E 6BT, United Kingdom}

\author{Lukas Graf}
\email{lukas.graf.14@ucl.ac.uk}
\affiliation{Department of Physics and Astronomy, University College London, London WC1E 6BT, United Kingdom}

\author{Suchita Kulkarni}
\email{suchita.kulkarni@oeaw.ac.at}
\affiliation{Institut f\"ur Hochenergiephysik, \"Osterreichische Akademie der Wissenschaften, Nikolsdorfer Gasse 18, A-1050 Wien, Austria}

\author{\hspace*{0cm}Sudhanwa\hspace*{1cm} \hspace*{-1cm} Patra}
\email{sudhanwa@mpi-hd.mpg.de}
\affiliation{Max-Planck-Institut f\"ur Kernphysik, Saupfercheckweg 1, 69117 Heidelberg, Germany}
\affiliation{Center of Excellence in Theoretical and Mathematical Sciences, Siksha \textquoteleft O\textquoteright\, Anusandhan University, Bhubaneswar-751030, India}

\author{Werner Rodejohann}
\email{werner.rodejohann@mpi-hd.mpg.de}
\affiliation{Max-Planck-Institut f\"ur Kernphysik, Saupfercheckweg 1, 69117 Heidelberg, Germany}

\author{Narendra Sahu}
\email{nsahu@iith.ac.in}
\affiliation{Department of Physics, Indian Institute of Technology, Hyderabad, Yeddumailaram, 502205, Telengana, India}

\author{Utpal Sarkar}
\email{utpal@prl.res.in}
\affiliation{Physical Research Laboratory, Ahmedabad 380 009, India}

\begin{abstract}
We interpret the 2 TeV excesses at the LHC in a left-right symmetric model with Higgs doublets and spontaneous 
$D$-parity violation. The light neutrino masses are understood via a linear seesaw, suppressed by a high $D$-parity 
breaking scale, and the heavy neutrinos have a pseudo-Dirac character. In addition, with a suppressed right-handed gauge 
coupling $g_R / g_L \approx 0.6$ in an $SO(10)$ embedding, we can thereby interpret the observed $eejj$ excess 
at CMS. We show that it can be reconciled with the diboson and dijet excesses within a simplified 
scenario based on our model. Moreover, we find that the mixing between the light and heavy neutrinos can be potentially large 
which would induce dominant non-standard contributions to neutrinoless double beta decay via long-range $\lambda$ 
and $\eta$ neutrino exchange.
\end{abstract}


\maketitle

\section{Introduction}

The origin of small neutrino mass, as confirmed by oscillation experiments~\cite{oscillation_expts}, is still a 
mystery in particle physics. Seesaw mechanisms are the leading candidates among beyond the Standard Model (SM) scenarios 
to explain sub-eV masses of the neutrinos. In the conventional type-I seesaw~\cite{typeI_seesaw}, the SM is extended by three 
$SU(2)_L$ singlet right handed neutrinos $\nu_R$ with hypercharge $Y=0$, while in case of type-II seesaw~\cite{typeII_seesaw} 
one adds a scalar triplet with hypercharge $Y=2$ to the SM. In either case, the neutrino masses can be given by $m_\nu = y^2 \langle H \rangle^2 
 / M$, where $M$ is the seesaw scale set by the masses of new particles and $y$ is an effective coupling. An important point to be noted in 
this scenario is that the new particles are {\it ad-hoc} and their masses are not controlled by the SM gauge symmetry $SU(2)_L \times U(1)_Y$. 
Left-right symmetric models~\cite{LR_model}, on the other hand, are extensions of the SM, where both type-I and type-II seesaw for 
sub-eV neutrino masses emerge naturally. The breaking of left-right symmetry then fixes the scale of seesaw $M$. By embedding the left-right 
symmetric model in a non-supersymmetric grand unified theory (GUT) one can find that $M$ can vary from TeV scale to a sub-GUT scale depending 
on the particle content and the pattern of symmetry breaking. While the generic solution $M \approx 10^{14}$~GeV and $y\approx 1$ will result 
in the observed small neutrino masses, models with relatively light $M \approx 100$~GeV - 1~TeV have the benefit that they can be probed 
directly at colliders such as the LHC. While this seems to require tiny couplings $y$, extending the heavy sterile neutrino sector or 
taking into account the flavour structure will allow the couplings and masses to conspire to produce small neutrino masses with large 
couplings $y$. Rather we discuss here the linear seesaw mechanism~\cite{linear_seesaw} and its possible implications to neutrino masses, 
lepton flavor violation, neutrinoless double beta decay and Collider studies in a class of TeV scale left-right symmetric model (LRSM). 

In this paper we propose an extended left-right symmetric model (LRSM) based on the gauge group $SU(2)_L \times SU(2)_R 
\times U(1)_{B-L}\times D$ with doublet Higgs and spontaneous $D$-parity breaking~\cite{Dparity-papers}, to explain the light 
neutrino masses via a linear seesaw mechanism. The extension of the LRSM~\cite{LR_model} is provided 
by a singlet fermion $S$. We show that the suppression of neutrino mass arises through the large $D$-parity breaking scale 
that generates the small scale of lepton number violation, while allowing $SU(2)_R\times U(1)_{B-L}$ to be broken at the TeV scale. 
We embed the low energy left-right symmetric model in a non-supersymmetric $SO(10)$ GUT to calculate the difference between 
the left and right sector gauge couplings $g_L$ and $g_R$ after $D$-parity breaking. 

The main focus of this paper is on the interpretation of the various excesses observed at the LHC around the energy 
scale of 2~TeV. Starting with the excess in two leptons, two jet $e e jj$ final state at CMS~\cite{cms_excess}, which can be better 
understood in our model due to suppressed right-handed $SU(2)_R$ gauge coupling at the electroweak (EW) scale. In addition, the fact 
that the heavy neutrinos are of Pseudo-Dirac type guarantees that the two leptons in the process have opposite sign, in agreement 
with the observation (as has also been noted in Ref.~\cite{Gluza:2015goa}). 

Starting with the excess in two leptons, two jet $e e jj$ final state at CMS~\cite{cms_excess}, which can be better 
understood in our model due to suppressed right-handed $SU(2)_R$ gauge coupling at the electroweak (EW) scale. In addition, the fact 
that the heavy neutrinos are of Pseudo-Dirac type guarantees that the two leptons in the process have opposite sign, in agreement 
with the observation (as has also been noted in Ref.~\cite{Gluza:2015goa}). 

Although the observation of the $eejj$ excess was one of the first motivations to consider LRSMs, subsequent reports of several 
resonance excesses in different final states with a mass around 2 TeV has led to an increased interest in these scenarios. 
Among them, most notable are, the diboson excess~\cite{Aad:2015owa} in fully hadronic decays of the final state bosons 
as reported by ATLAS, the diboson excess~\cite{Khachatryan:2014gha} in semi-leptonic final state observed by CMS and 
the excess in W boson, SM Higgs final state~\cite{CMS-PAS-EXO-14-010} reported also by CMS. For any LRSM, explaining 
the $eejj$ excess, it is important to have consistency with all these excesses as well. It is crucial to note that 
none of these excesses are statistically significant and the results of ongoing Run - 2 at 13 TeV are necessary for 
the confirmation. We take into account the above said and few other channels to which the model under consideration 
is sensitive and perform a rough simultaneous fit to the mass of the heavy neutrino, 
the right handed $W$ boson along with their respective mixing angles.

The paper is arranged as follows. In Section II, we briefly discuss the LRSM with doublet Higgs and spontaneous 
$D$-parity breaking. Section III is devoted to the calculation of the neutrino mass via the linear seesaw mechanism 
and the consequences for neutrinoless double beta decay. Section IV estimates the mismatch 
between the gauge couplings by embedding the low energy LRSM in a non-supersymmetric SO(10) GUT. In Section V we then 
discuss how the various excesses at the LHC can be understood in our framework before we conclude in Section VI.

\section{LRSM with Higgs doublets}
In the usual left-right symmetric extension~\cite{LR_model} of the Standard Model, the gauge group is expanded to $SU(3)_c 
\times SU(2)_L \times SU(2)_R\times U(1)_{B-L} \equiv {\cal G}_{LR}$, where $B$ is baryon number and $L$ is lepton number. 
The electric charge $Q$ and the hypercharge $Y$ are related to the quantum numbers of the group as~\cite{LR_model}
\begin{equation}
	Q = T_{3L} + T_{3R} + \frac{B-L}{2} = T_{3L} + Y \,.
\end{equation}
The conservation of left-right parity implies that all left-handed fermions have right-handed partners. Therefore, the model 
accommodates automatically a right-handed neutrino $N$ per family, which is singlet under the SM gauge group.

In the LRSM,the choice of the discrete left-right symmetry i.e. either Parity $\mathcal{P}$ or charge conjugation symmetry $\mathcal{C}$, plays a crucial role. In addition to the usual parity of the Lorentz group (denoted by ${\cal P}$), a discrete left-right symmetry called 
$D$-parity (similar to charge conjugation symmetry) that acts on the left-right symmetric gauge group is also assumed 
to be conserved. For the usual fermions, i.e. quarks and leptons, these two parity operations can be identified with 
each other. However, scalar particles, which transform under the Lorentz group trivially, can be transformed under 
the $D$-parity non-trivially, as they belong to non-trivial representations of the left-right symmetric group. 
For example, in the usual convention, the charge conjugation operator of the left-right symmetric group is identified 
with $D$-parity. Then the $D$-parity would transform a scalar belonging to a representation ${\cal R}$ of the left-right 
symmetric group to a representation ${\cal R}^\ast$. In that case the breaking of $D$-parity could take place at a different 
scale than the breaking of parity ${\cal P}$ of the Lorentz group. Since the breaking of the left-right symmetric group 
is always associated with the breaking of parity ${\cal P}$ of the Lorentz group, this means that the $D$-parity breaking 
scale could be decoupled from the left-right symmetry breaking scale. Details of left-right symmetric models with 
spontaneous $D$-parity breaking can be found in~\cite{Dparity-papers}.

The particle content of the model and the gauge transformation under ${\cal G}_{LR} \times D$ is given as follows \cite{potential_min}. The fermionic 
representations are (per generation)
\begin{gather}
	Q_{L}=
		\begin{pmatrix}
			u_{L} \\
			d_{L}
		\end{pmatrix}  \equiv (3,2,1,1/3),  
	Q_{R}=
		\begin{pmatrix}
			u_{R}\\
			d_{R}
		\end{pmatrix} \equiv (3,1,2,1/3)\,, \nonumber \\	  
	\ell_{L}=
		\begin{pmatrix}
			\nu_{L}\\
			e_{L}
		\end{pmatrix} \equiv (1,2,1,-1),~ 
	\ell_{R}=
		\begin{pmatrix} 
			N\\
			e_{R}
		\end{pmatrix} \equiv (1,1,2,-1) \,, \nonumber\\	  
	S \equiv (1,1,1,0) \,.
\end{gather}
 The scalar sector consists of
\begin{gather}
	H_L= 
		\begin{pmatrix} 
			h_L^+\\
			h_L^0
		\end{pmatrix}\equiv (1,2,1,1),
	H_R=
		\begin{pmatrix} 
			h_R^+  \\
			h_R^0
		\end{pmatrix} \equiv (1,1,2,1), \nonumber \\
	\Phi =
		\begin{pmatrix} 
			\phi_{1}^0     &  \phi_{2}^+ \\
			\phi_{1}^-     &  \phi_{2}^0
		\end{pmatrix} \equiv (1,2,2,0), \quad 
	\eta \equiv (1,1,1,0)  \,,
\end{gather}
where the quantum numbers inside the parentheses are under the gauge group $SU(3)_c \times SU(2)_L \times SU(2)_R\times 
U(1)_{B-L}$. The only particle $\eta$ is odd under $D$-parity while rest of the particles are even. In the fermion sector 
we have included singlet fermions $S$, which will eventually couple to the left-handed neutrinos.

\subsection{Symmetry breaking}
The electroweak symmetry is broken by a bi-doublet scalar field $\Phi$, which gives Dirac masses to all charged fermions 
and also to neutrinos. In addition, we introduce a right-handed doublet scalar field $H_R$ and its left-right symmetric 
partner $H_L$. The former breaks the left-right symmetry and mixes the singlet fermion $S$ with the right-handed neutrinos, 
while the field $H_L$ acquires a much smaller $vev$ and mixes $S$ with the left-handed neutrinos. Finally we add a $D$-parity 
odd singlet scalar field $\eta$, whose $vev$ breaks the $D$-parity while keeping the left-right gauge symmetry intact. 
The symmetry breaking pattern can thus be written as \cite{Dparity-papers,potential_min}
\begin{gather}
\label{symmetrybreaking} 
	SU(3)_c \times SU(2)_L \times SU(2)_R \times U(1)_{B-L} \times D 	\nonumber\\
		\downarrow \langle \eta \rangle 																\nonumber\\
	SU(3)_c \times SU(2)_L \times SU(2)_R \times U(1)_{B-L}   				\nonumber\\
		\downarrow \langle H_R \rangle																	\nonumber\\
	SU(3)_c \times SU(2)_L \times U(1)_Y 															\nonumber\\
		\downarrow \langle \Phi \rangle 																\nonumber\\
	SU(3)_c \times U(1)_{em}\,.
\end{gather}
We define the $vev$s of the various fields as
\begin{align} 
	\langle \Phi \rangle = k_1, k_2, 	\,\, 
	\langle H_L \rangle  = v_L,				\,\,
	\langle H_R \rangle  = v_R, 			\,\,  
	\langle \eta \rangle = \eta_P\,.
\end{align}
The scalar potential contains the usual quadratic and quartic terms. Several phenomenology aspects of this potential were previously discussed in~\cite{potential_min}. 
The terms relevant for the purpose of our discussion are given by
\begin{align}
	V&\supset 
			 \mu_h^2(H_L^\dagger H_L + H_R^\dagger H_R) \nonumber \\
		&+ \mu_1(H_L^\dagger \Phi H_R + H_R^\dagger \Phi^\dagger H_L) \nonumber \\
		&+ \mu_2(H_L^\dagger \tilde{\Phi} H_R + H_R^\dagger \tilde{\Phi}^\dagger H_L) \nonumber \\
		&+ M^\prime \eta(H_L^\dagger H_L - H_R^\dagger H_R),
		\label{eq:6}
\end{align}
with $\tilde{\Phi} = \tau_2 \Phi^*_1 \tau_2$ while $\mu_h$ as the bare mass term for scalar doublets 
and $\mu_{1,2}, M^\prime$ being the trilinear coupling parameters having mass dimension one. Minimization 
of this potential with respect to the various fields would give us the consistency conditions and relationships 
between the various $vev$s \cite{potential_min}. The minimization condition that is relevant to our discussions 
is given as
\begin{gather}
	v_R\frac{\partial V}{\partial v_L}-v_L \frac{\partial V}{\partial v_R} = 0 \nonumber \\
	\Rightarrow 2 M^\prime \eta_P v_L v_R- (\mu_1 k_1 + \mu_2 k_2)(v_R^2 -v_L^2) = 0.
\end{gather}
From the above expression it is evident that the minimum allows a left-right symmetric solution only 
for $\eta_P=0$. When the $D$-parity is broken by the $vev$ of the $D$-parity odd singlet scalar 
$\langle \eta \rangle = \eta_P$, the $vev$ of $H_L$ becomes much smaller than the $vev$ of $H_R$. 
Defining $M_D$ as the $D$-parity breaking scale and using allowed model parameters, 
$$ 
	\eta_P \simeq M^\prime \simeq M_D\,, ~~~~ v_R > k_1, k_2 \sim M_W\,,  
$$
we obtain
\begin{equation}
\label{vL-equation}
	v_L \simeq \frac{- (\mu_1 k_1 + \mu_2 k_2) v_R}{ M^\prime \eta_P}\,,
\end{equation}
which allows $\eta_P \gg v_R \gg v_L$. The suppression of the $vev$ $v_L$ by the $D$-parity breaking scale 
is what suppresses the neutrino mass. A representative set of parameters would be $\eta_P ={\cal O}(10^{9})$~GeV, 
$v_R = {\cal O}(1)$~TeV and $v_L = {\cal O}(10^{-9})$~GeV which gives rise to correct neutrino masses.

Due to the spontaneous $D$-parity breaking, the effective masses of the left-handed and the right-handed 
doublet scalar fields $H_L$ and $H_R$ are obtained from Eq.~(\ref{eq:6}) as
\begin{align}
\label{mr}
	\mu^2_L = \mu^2_h +M^\prime \eta_P \,,\,\,\,\,\mu^2_R = \mu^2_h -M^\prime \eta_P \,.
\end{align}
Similarly to the $D$-parity violating conditions in models with triplet Higgs scalars~\cite{Dparity-papers}, 
a fine-tuning of the parameter $M$ for a given value of the $vev$ of the singlet field $\eta_P$ can 
allow a right-handed doublet field $H_R$ with mass in the TeV range so as to make it accessible at the LHC. 
At the same time, the mass of the left-handed scalar doublet $H_L$ can be several orders of magnitude larger, 
which then implies that the $vev$ of this field must be orders of magnitude smaller than the electroweak symmetry 
breaking scale. In this work we will assume that the Higgs fields are heavier than the right-handed $W_R$ boson 
to be discussed below. This will simplify the discussion of the LHC signatures as the only non-SM particle assumed 
lighter than $W_R$ will be a heavy neutrino.

\subsection{Gauge bosons}
After spontaneous symmetry breaking of the left-right gauge and the $D$-parity the mass matrix 
for the left- and right-handed charged gauge bosons $W_L, W_R$ is~\cite{potential_min}
\begin{align}
	M_{W^{\pm}}^2 &= \frac{1}{4}
	\left(
		\begin{array}{c|cc}&W^{+}_L&W^{+}_R \\
			\hline
			W^{-}_L&g^2_L \left(k^2+v_L^2\right)&- 2g_L g_R k^*_1 k_2\\
			W_R^{-}&- 2g_L g_R k_1 k^*_2 & g^2_R \left(k^2+v_R^2\right) 
		\end{array}
	\right) ,
	\label{eq:10}
\end{align}
where $k^2 = |k_1|^2 + |k_2|^2$. Diagonalizing the above mass matrix, the physical masses 
of the charged gauge bosons are
\begin{align}
	M_W^2 &\approx \frac{1}{4} g_L^2 \left(|k_1|^2 + |k_2|^2\right)\, ,\nonumber \\
	M_{W_R}^2 &\approx \frac{1}{4} g_R^2 v_R^2 \,,
\end{align}
where $g_L$ ($g_R$) is the gauge coupling for gauge group $SU(2)_L$ ($SU(2)_R$). 
The mixing between the left-handed and right-handed gauge bosons which can be obtained 
from Eq.~(\ref{eq:10}) is
\begin{equation}
	|\sin\theta_\text{LR}^W| 
		\approx 2 \frac{g_R}{g_L}\frac{M^2_W}{M^2_{W_R}} \frac{k_2/k_1}{1+k^2_2/k^2_1} \,.
\end{equation}
Choosing a hierarchy between the $vev$s of the bi-doublet can suppress this mixing, but a mild hierarchy 
is phenomenologically consistent with fermion masses with large hierarchy between respective Yukawa couplings. 
With $g_R / g_L \approx 0.6$ from the embedding in SO(10) (cf. Section IV) one can get the $W_L-W_R$ mixing 
around $10^{-3}$ with a $W_R$ mass around 2~TeV as suggested by the diboson excess (cf. Section V).

Similarly, we can write down the mass-squared matrix for the neutral gauge bosons $W_L^3$, $W_R^3$ and $B$ as~\cite{potential_min}
\begin{equation}
\label{neumass}
	M^2 =\frac{1}{4}
	\left(\begin{array}{ccc}
		g^2_L (k^2+v_L^2) & -g_L g_R k^2      & -g_L g_{BL} v_L^2     \\
		-g_L g_R k^2      & g^2_R (k^2+v_R^2) & -g_R g_{BL}  v_R^2    \\
		-g_L g_{BL} v_L^2 & -g_R g_{BL} v_R^2 & g_{BL}^2(v_L^2+v_R^2)
	\end{array}\right)
\end{equation}
where $g_{BL}$ is the gauge coupling for $U(1)_{B-L}$ which is related to 
the $U(1)_Y$ SM gauge coupling as $1/g^2_Y = 1/g^2_R + 1/g^2_{BL}$.
It is clear from the above mass matrix that one of the mass eigenvalues is zero as the determinant 
vanishes and we denote this physical state as photon. The diagonalization procedure yields SM $Z$ 
and heavy $Z_R$ mass eigenstates
\begin{align}
	M_Z^2 &\approx \frac{1}{2} \frac{g_L^2}{c_W^2} (k^2 + v_L^2)
		- \frac{g_L^2}{2c_W^2 v_R^2} \left( c_M^2 k^2 
		- s_M^2 v_L^2\right)^2, \nonumber \\     
	M_{Z_R}^2 &\approx \frac{1}{2} (g_{BL}^2 + g_R^2 ) v_R^2
		+ \frac{1}{2} \left(g_R^2 c_M^2 k^2 
		+ g_{BL}^2 s_M^2 v_L^2\right), 
\end{align}
where $s_M \equiv \sin\theta_M = g_{BL} / \sqrt{g_{BL}^2 + g_R^2}, c_M=\sqrt{1-s^2_M}$. 
As we have a hierarchy between the different $vev$s according to $v_R \gg k_1, k_2\gg v_L$, 
one can expand the heavy gauge boson masses in terms of $1/v_R^2$ as
\begin{align}
	M_{W_R}^2 &\approx \frac{1}{2} g_R^2 v_R^2
		\left(1 + \frac{k^2}{v_R^2} \right), \nonumber\\
	M_{Z_R}^2 &\approx \frac{1}{2} (g_{BL}^2 + g_R^2) v_R^2 
		\left(1 + \frac{c_M^4 k^2 + s_M^4 v_L^2}{v_R^2}\right)\, .
\end{align}
For generic $g_R \neq g_L$ scenarios, the mass relation between the two gauge bosons is given by
\begin{align}
	\frac{M_{Z_R}}{M_{W_R}} = \frac{\sqrt{2} g_R/g_L}{\sqrt{(g_R/g_L)^2-\tan^2\theta_W}}\,,
\end{align}
with the SM weak angle $\theta_W$. Hence, $M_{Z_R} > M_{W_R}$ and numerically for $g_R/g_L = 0.6$ 
we have $M_{Z_R} \approx 4 M_{W_R}$. Using the above relation one can derive a lower bound $g_R/g_L > 0.56$. 
Using 8 TeV ATLAS $20.3$ fb$^{-1}$ luminosity data, the derived bound on $Z_R$ is $M_{Z_R} > 2.2$~TeV \cite{ZR-bound}
for a range of $g_R$ values consistent with the above mass relation. 
It is interesting to note that the ratio of couplings needed to reconcile the excesses in our model 
is close to the theoretical limit, making the mass of the right-handed $Z$ boson arbitrarily heavier 
as compared to $W_R$. This means, that the $Z_R$ will be potentially out of reach at the LHC.

\section{Neutrino mass via Linear seesaw}
Let us write down the relevant Yukawa terms in the Lagrangian that contribute to the fermion masses~\cite{potential_min},
\begin{align}
\label{Yuk-Lag}
	\mathcal{L}_{\rm Yuk} &= 
		h_{\ell} \overline{\ell_R} \Phi \ell_L  
		+ \widetilde{h_\ell} \overline{\ell_R} \widetilde{\Phi} \ell_L  
		+ f_{R} \overline{S} \widetilde{H_R} \ell_R  \nonumber\\
	  &+ f_{L} \overline{S} \widetilde{H_L} \ell_L 
		+  \mu_S \overline{(S_L)^c} S_L 
		+ \text{h.c.}, 
\end{align}
with $\widetilde{H_j} = i\tau_2 H_j^*$  ($j=L, R$) and $\widetilde{\Phi}=\tau_2 \Phi^* \tau_2$. The singlet Majorana 
field $S$ in Eq.\ (\ref{Yuk-Lag}) is defined as
\begin{equation}
	S=(S_L + (S_L)^c) / \sqrt{2}\,.
\end{equation}
The $vev$s of $H_L$, $H_R$ and $\Phi$ will generate a mass matrix of the neutral fermions in the Majorana basis 
$\psi_L^T=(\nu_L, (\nu^c)_L, S_L)$ as
\begin{eqnarray}
	{\cal L}_{\rm mass}  = 
	\begin{pmatrix} 
		\overline{(\nu_L)^c} \, \overline{N} \,  \overline{(S_L)^c}
	\end{pmatrix}
  M_N
	\begin{pmatrix}
		\nu_L \\ (\nu^c)_L \\ S_L
	\end{pmatrix} ,\nonumber
\end{eqnarray}
with the neutral lepton mass matrix 
\begin{eqnarray}
	M_N  &=& 
	\begin{pmatrix}
		0              & h_\ell k_1+\widetilde{h_\ell}k_2          & f_L v_L\\
		h_\ell^T k_1+\widetilde{h_\ell}^T k_2        & 0           & f_R v_R \\
		f^T_L v_L      & f^T_R v_R   & \mu_S
	\end{pmatrix} \nonumber\\
  &\equiv&  
  \begin{pmatrix}
    0              & m_D         & m_L\\
		m^T_D        & 0           & M \\
		m^T_L      & M^T   & \mu_S
	\end{pmatrix}.
\label{eq:mtot}       
\end{eqnarray}
From the interaction Lagrangian given in Eq.~(\ref{Yuk-Lag}), $m_D = h_\ell k_1+\widetilde{h_\ell}k_2$ is the Dirac neutrino 
mass mixing for the left- and right-handed neutrino states, $M = f_R \langle H_R \rangle = f_R v_R$ is the heaviest Dirac 
neutrino mass term mixing $N$-$S$ while $m_L = f_L \langle H_L \rangle = f_L v_L$ is the small lepton number violating term 
arising from the induced $vev$ of $H_L$. 

It is usually difficult to get a small mass $\mu_S$ for the singlet field $S$ in any left-right symmetric model, particularly 
the ones originating from a Grand Unified Theory (GUT). In the present model, we prevent the bare mass of the field $S$ by 
introducing a global $U(1)_X$ symmetry, and assigning $X = 1$ for the field $S$ and $X = -1$ for the fields $H_L$ and $H_R$. 
This ensures $\mu_S= 0$ while allowing all other terms in the mass matrix. Although the $vev$ of the field $H_R$ breaks the 
$U(1)_X$ global symmetry, this cannot generate $\mu_S$ as there are no $D$-parity even singlet scalar fields acquiring 
$vev$s in the model. The $D$-parity odd field $\eta$ does not have any interaction with the field $S$, and hence, its $vev$ 
cannot contribute to the mass of $S$. Therefore, the neutrino mass is entirely given by a linear seesaw. From 
Eqs.~\eqref{vL-equation}, \eqref{eq:mtot} and with the hierarchy $M > m_D \gg m_L$ we obtain the light neutrino 
mass matrix as
\begin{align}
\label{eq:mnu}
	m_\nu &= \frac{\mu_1 k_1 + \mu_2 k_2}{M'\eta_P}
		\left[ m^T_D \left(f_R^{-1} f_L \right) 
		\!\!+\!\! \left( f^T_L {f_R^{-1}}^T\right)m_D\right] \nonumber\\
		&= m^T_D\, M^{-1} m_L \mbox{+\,transpose}\, . 
\end{align}
Additionally, we get two heavy pseudo-Dirac states, whose masses are separated by the light neutrino mass, given by 
\begin{equation}
	\widetilde{M} \approx \pm M + m_{\nu}\,.
\end{equation} 
From Eq.~\eqref{eq:mnu} it is clear that the light neutrino mass is suppressed by the parity breaking scale. 
The smallness of $v_L$ thus ensures the smallness of the observed sub-eV scale neutrino masses. The $SU(2)_R\times U(1)_{B-L}$ 
breaking scale $v_R$ can be as low as a few TeV. This is in contrast to the usual left-right symmetric model without 
$D$-parity, where the neutrino mass is suppressed by $v_R$ and hence cannot be brought to TeV scales easily.    

\begin{figure}[t]
\centering
\includegraphics[width=0.7\linewidth]{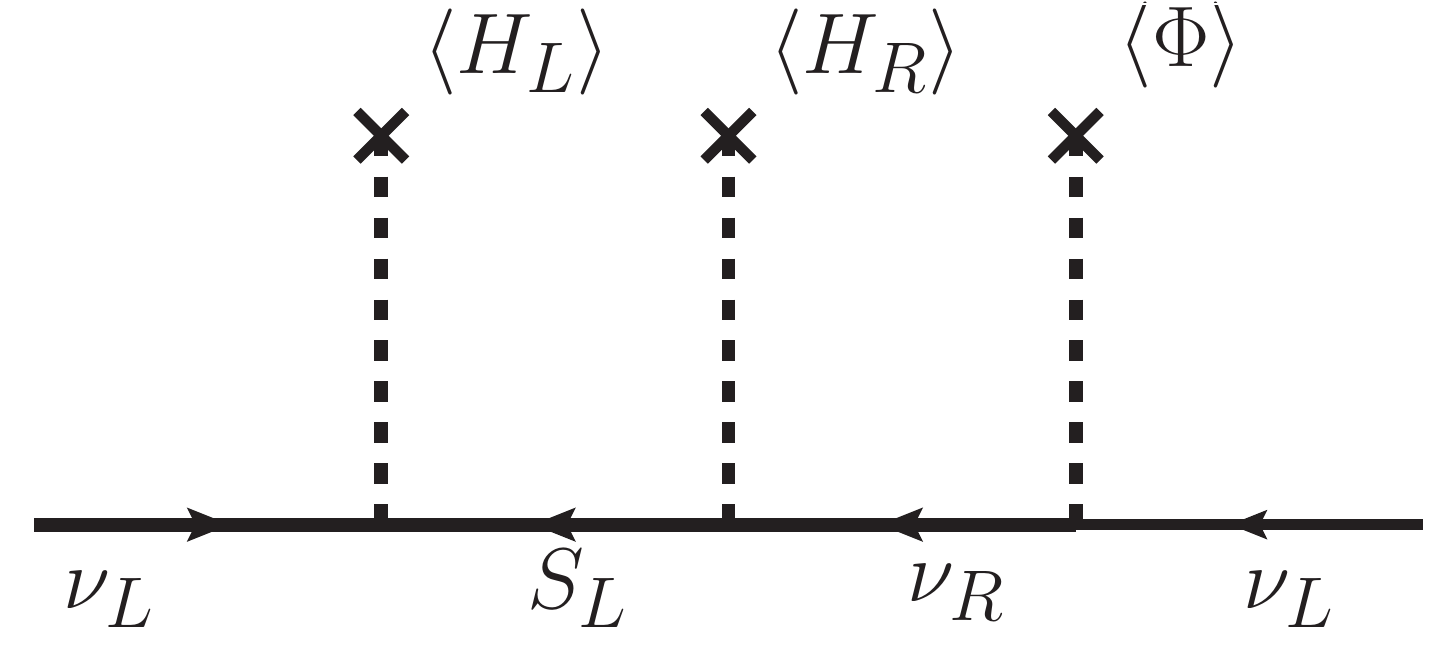}
\caption{Majorana mass of the light neutrinos arising through the combined $vev$s of $H_L$ and $H_R$.}
\label{fig:neutrino-mass}
\end{figure}

Note that the $vev$s of $H_L$ and $H_R$ break lepton number by one unit individually; thus, their combination 
gives the Majorana nature of light neutrinos. This can also be understood from the Feynman diagram as shown 
in Fig.~\ref{fig:neutrino-mass}.

\subsection{Unitarity and lepton flavor violation}
Order-of-magnitude-wise and neglecting any flavour structure, we have $m_D \sim 1$~GeV, $M \sim 10^3$ GeV 
and $m_L \sim 100$~eV, resulting in $m_\nu$ of order 0.1~eV. As usual, seesaw mechanisms with additional fermions 
induce violation of unitarity \cite{hh}. Defining $\Theta$, as the mixing between light and heavy neutrinos, 
and $\epsilon$, as a measure of deviation from unitarity in the PMNS mixing matrix in the light neutrino sector, 
one can express the leptonic mixing matrix as $N = (1 + \epsilon) U^\text{PMNS}$ with \cite{Antusch:2014woa}
\begin{equation}
	\epsilon \simeq -2 \, m_D  (M^T)^{-1} \, (M^\ast)^{-1} m_D^\dagger 
	\equiv -2 \Theta \, \Theta^\dagger\,. 
\end{equation}
The unitarity violation is therefore of the order $m_D^2/M^2$. The off-diagonal $e\mu$, $e\tau$ and $\mu\tau$ 
elements of $\epsilon$ are currently constrained to be smaller than $10^{-5}$, $10^{-4}$ and $4\times 10^{-4}$, 
respectively \cite{Antusch:2014woa}. In particular the $e\mu$ entry plays a leading role, as it induces the decay 
$\mu \to e\gamma$. In fact, the branching ratio is given by \cite{Ibarra:2011xn}
\begin{align}
	{\rm BR}(\mu \to e \gamma) = 
		\frac{3\alpha}{32\pi} \sum_{i=1}^3 f\left(\frac{M_i}{M_W}\right) 
		\left| \Theta_{\mu i}^\ast \, \Theta _{e i}\right|^2 \,,
\end{align}
where $M_i$ are the heavy neutral fermion masses and $f(M^2_i/M^2_W)$ is a loop-function of order one. We do not 
study in detail here the flavor structure of the Yukawa matrices, but setting $M_i = 1$~TeV would result in
\begin{align}
	{\rm BR}(\mu \to e \gamma) \simeq 
		8.4 \times 10^{-14} \cdot
		\left(\frac{|(\Theta\Theta^\dagger)_{e\mu}|}{10^{-5}}\right)^2\,.
\end{align}
The current limit derived from MEG experiment is ${\rm BR}(\mu \to e \gamma) < 5.7 \times 10^{-13}$ \cite{MEG} giving 
a similar limit on $\Theta \sim m_D/M$ as the unitarity constraints. 

Since we work in a left-right symmetric framework, there are several additional diagrams 
that contribute to low energy lepton flavor violating processes. In the present doublet version 
of the left-right symmetric model, the most relevant contribution to $\mu\to e\gamma$ is from the loop exchange 
of a heavy right-handed $W_R$ boson and neutrino $N$. With respect to the LHC signatures we will assume that 
there is only one $N$ that dominantly couples to electrons, and that the neutrino generations are aligned with 
the charged lepton flavors. In this scenario we neglect the $W_R$ induced flavor violation.

\subsection{Neutrinoless double beta decay}
LRSMs can give rise to a large number of non-standard contributions \cite{0vbb,0nubb-LRa,0nubb-LRb,0nubb-LRc,0nubb-spatra-jhep, Simkovic:2015} to neutrinoless double beta decay in addition 
to the standard light neutrino exchange neutrinos. Regarding the heavy neutrinos, as they form three quasi-Dirac 
pairs of mass $M$ with essentially negligible splitting of order $m_\nu$, i.e. of order $10^{-10}$~GeV, their  
contributions via left- or right-handed currents are negligible. 
However, the active neutrinos can couple to the right-handed $W_R$ bosons through the LR neutrino mixing of order 
$\Theta \approx m_D/M$. Such contributions are a priori not negligible and potentially interesting. Diagrams exist 
in which a light neutrino mediates neutrinoless double beta decay by coupling to both the left- and right-handed 
$W$ boson. These contributions, referred to as $\lambda$ and $\eta$ diagrams \cite{0vbb}, 
are not suppressed 
by the light Majorana neutrino masses, as no helicity flip is required. The former is suppressed by $(W/W_R)^2$ 
in the amplitude as the one of the Fermi interactions in the standard exchange is replaced by a right-handed current, 
whereas the latter is suppressed by $\sin\theta_{LR}^W$ from a $W_L - W_R$ mixing. The neutrinoless double beta decay 
amplitudes for these two contributions are approximately given by~\cite{0vbb}
\begin{align}
	{\cal A}^{0\nu}_\lambda &\approx 
		\frac{10^{-2}}{|q|} \left(\frac{g_R}{g_L}\right)^2 \left(\frac{M_W}{M_{W_R}}\right)^2 
		(U^\text{PMNS} \Theta^\dagger)_{ee}, \nonumber\\
	{\cal A}^{0\nu}_\eta &\approx 
		\frac{1}{|q|} \left(\frac{g_R}{g_L}\right) \sin\theta_{LR}^W
		(U^\text{PMNS} \Theta^\dagger)_{ee}.
\end{align}
Here, $|q| \simeq 100$~MeV is the virtual neutrino momentum and the factor $10^{-2}$ comes 
from the nuclear matrix element as normalized with respect 
to the light neutrino exchange diagram (we assume Germanium as isotope).  Defining 
$\langle \theta \rangle^{0\nu} \equiv |(U^\text{PMNS} \Theta^\dagger)_{ee}|$, the half-life for neutrinoless 
double beta decay arising from the $\lambda$ and $\eta$ diagrams can be expressed as
\begin{align}
\label{eq:halflives}
	\frac{3\times 10^{25}\text{ y}}{T^{0\nu}_{\rm \lambda}} &\approx
		\left(\frac{g_R/g_L}{0.6}\right)^4
		\left(\frac{\langle \theta \rangle^{0\nu}}{10^{-3.0}}\right)^2
		\left(\frac{2\text{ TeV}}{M_{W_R}}\right)^4 \!, \nonumber\\
	\frac{3\times 10^{25}\text{ y}}{T^{0\nu}_{\rm \eta}} &\approx
		\left(\frac{g_R/g_L}{0.6}\right)^2
		\left(\frac{\langle \theta \rangle^{0\nu}}{10^{-5.3}}\right)^2
		\left(\frac{|\sin\theta_{LR}^W|}{10^{-3}}\right)^2 \!.
\end{align}
These predictions can be compared with the current limit $T^{0\nu} \gtrsim 3\times 10^{25}$~y on the half-life 
of neutrinoless double decay \cite{GERDA, KamLAND-Zen, EXO}. This means that the individual contributions would 
alone begin to saturate the current experimental limit for $M_{W_R} \approx 2$~TeV and $\sin\theta_{LR}^W \approx 10^{-3}$ 
as suggested by the LHC excesses (cf. section~\ref{sec:lhc}), as well as the potentially large LR neutrino mixing 
$\langle \theta \rangle^{0\nu} \approx 10^{-5} - 10^{-3}$ in the linear seesaw. In this conclusion, we neglect 
possible cancellations between contributions. In the regime suggested by the LHC excesses, the $\eta$~diagram contribution to neutrinoless double beta decay half-life gives the strongest constraint on $\langle \theta \rangle^{0\nu}$ and it thus sharply restricts the allowed mixing between the light and heavy neutrinos.

\section{Embedding in $SO(10)$ and $g_L/g_R$}
Our framework of a LRSM with doublet and bi-doublet Higgs fields can be embedded in a non-supersymmetric SO(10) 
model including gauge coupling unification and TeV scale $W_R$ gauge bosons. The breaking scheme, which has 
a Pati-Salam symmetry as an intermediate step can be as follows \cite{Dparity-papers}
\begin{align}
\label{eq:BreakingChain}
	SO(10) \mathop{\longrightarrow}^{M_U} \mathcal{G}_{224D} \,
  	      \mathop{\longrightarrow}^{M_D} \mathcal{G}_{224}  \,
   	      \mathop{\longrightarrow}^{M_C} \mathcal{G}_{2213} \, 
   	      \mathop{\longrightarrow}^{M_R} \mathcal{G}_{SM}   \,
   	      \mathop{\longrightarrow}^{M_Z} \mathcal{G}_{13}   \, \nonumber 
\end{align}
The spontaneous symmetry breaking of $SO(10)$ down to low energies is provided by different Higgs multiplets 
contained in various Higgs representations of $SO(10)$, i.e. $10_{H}$, $16_{H}$, $54_{H}$ and $210_{H}$. 
The first stage of symmetry breaking $SO(10) \to \mathcal{G}_{224D}$ with $g_{L} = g_{R}$ is achieved by 
assigning a non-zero $vev$ to $54$-dimensional representation of $SO(10)$, i.e., $\langle \rho(1, 1, 1)\rangle \in 54_H$. 
Since $\rho$ is a $D$-parity even scalar and singlet under the Pati-Salam gauge group, the latter remains intact 
even after $\rho$ taking its non-zero $vev$. The second stage of symmetry breaking $\mathcal{G}_{224D} \to \mathcal{G}_{224}$ 
happens when the $D$-parity odd scalar singlet $\langle \eta (1,1,1)\rangle \in 210_H$ acquires a non-zero $vev$ leading 
to $g_L \neq g_R$. The third stage of symmetry breaking $\mathcal{G}_{224} \to \mathcal{G}_{2213}$ occurs when 
$\Sigma(1,1,15) \in 210_H$ takes a non-zero $vev$ by breaking $SU(4)_C$ down to $SU(3)_c \times U(1)_{B-L}$ at 
the mass scale $M_C$. The subsequent step of symmetry breaking $\mathcal{G}_{2213}$ down to the SM gauge group 
$\mathcal{G}_{213}$ is achieved by assigning a non-zero $vev$ to Higgs doublet $H_R$ contained in $16_{H}$. 
The final stage of symmetry breaking is provided by the SM Higgs doublet contained in $10_H$ of $SO(10)$. 

The motivation for considering such a long chain of symmetry breaking of $SO(10)$ GUT down to the low energy 
theory $U(1)_{\rm em} \times SU(3)_c$ is as follows: the inclusion of the intermediate symmetry breaking step 
$\mathcal{G}_{224D}$ around $10^8-10^{10.5}$~GeV is to give a small $vev$ to $H_L$ so that the light neutrino 
masses can be explained via a linear seesaw mechanism. Moreover, the Pati-Salam symmetry with $D$-parity 
invariance at the highest intermediate scale ensures gauge coupling unification since above this scale 
$M_D$ ($D$-parity breaking scale), we have only two gauge couplings to evolve, i.e.\ $SU(2)_L \equiv SU(2)_R$ 
and $SU(4)_C$. The inclusion of the subsequent intermediate symmetry breaking $\mathcal{G}_{224}$ at $M_C$ 
is required, as this gives a possibility of explaining the baryon asymmetry of the universe via post-sphaleron 
baryogenesis \cite{postsp}. The diquark Higgs scalars, if additionally included in the framework, also get 
their masses at $M_C$ ($\approx$ a few TeV), 
leading 
to neutron-antineutron oscillation with mixing time close to ongoing experimental search limits 
\cite{nnbarrev, nnbarrev1, postsp}. The crucial breaking step $\mathcal{G}_{2213} \to \mathcal{G}_{SM}\,$ 
occurs at the $M_R$ scale such that one can have $M_{W_R}$ within the TeV range. At the end, the SM symmetry 
breaking via a bi-doublet is needed for providing correct masses and mixing of fermions as well as to give 
mass to gauge bosons and the Higgs scalar.

\begin{figure}[t!]
\centering
\includegraphics[width=1.0\linewidth]{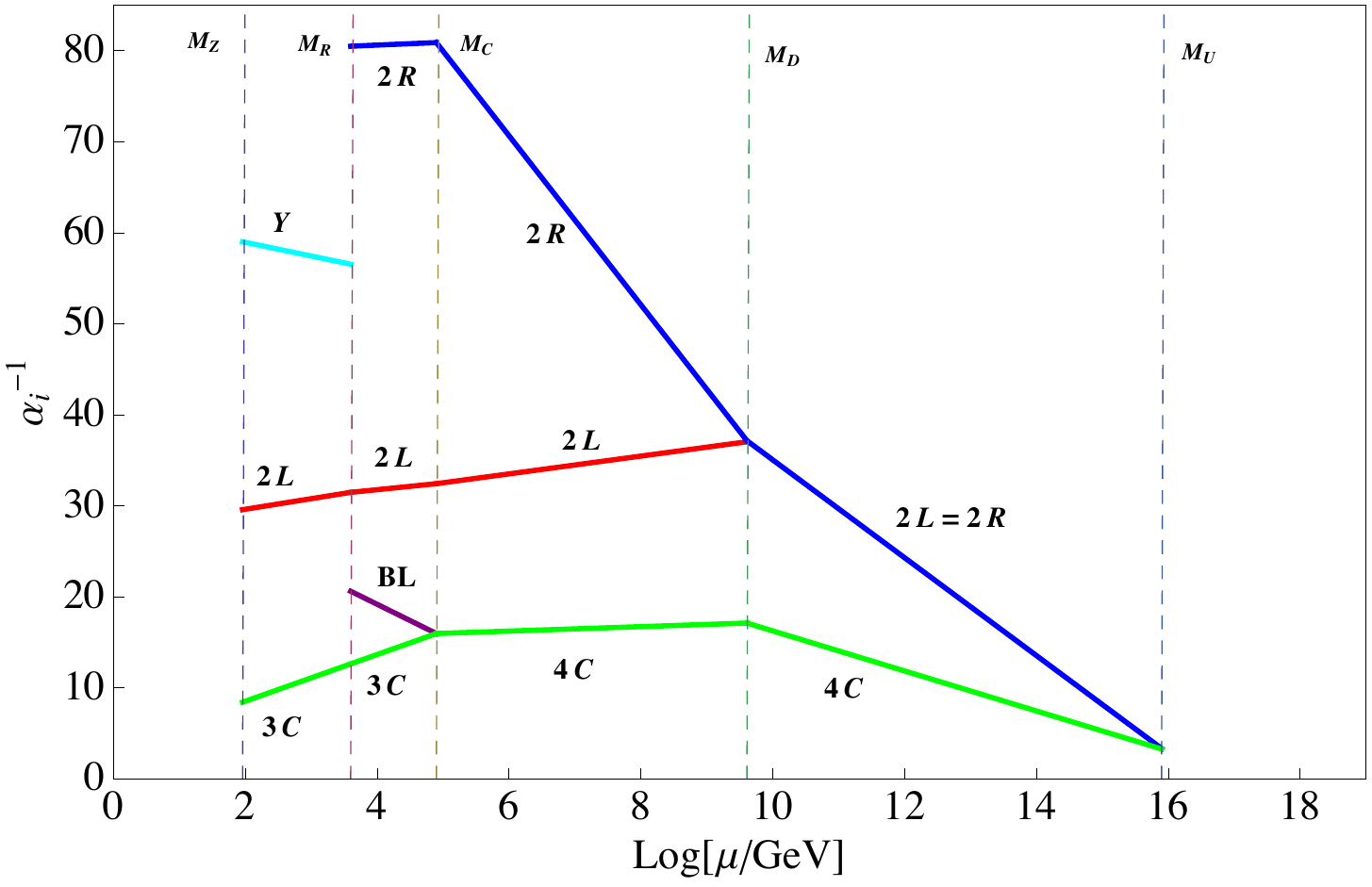}
\caption{One loop renormalization group evolution of gauge couplings with Pati-Salam symmetry $G_{224D}$ 
as the highest intermediate symmetry breaking.}
\label{fig:RG-evolution}
\end{figure}
Figure~\ref{fig:RG-evolution} shows the gauge coupling unification, with the associated mass scales 
and the prediction for the coupling ratio between the $SU(2)_L$ and $SU(2)_R$ gauge groups near the electroweak 
scale. However, we have introduced an extra $16_{H^\prime}$ and $210_{H^\prime}$ in addition to the usual 
$10_{H}$, $16_{H}$, $54_{H}$ and $210_{H}$ while performing the renormalization group evolution in order 
to achieve the unification of gauge couplings. For a detailed discussion on the relevant formalism, 
see Refs.~\cite{0nubb-spatra-jhep, nnbarrev1}. From Fig.~\ref{fig:RG-evolution}, the various mass 
scales are found to be $M_R = 8$~TeV, $M_C = 10^5$~GeV, $M_D = 10^{9.6}$~GeV and $M_U = 10^{15.9}$~GeV, 
which are consistent with the gauge coupling unification while predicting the desired coupling ratio 
between $g_L$ and $g_R$ at TeV scale,
\begin{align}
	\frac{g_R}{g_L} \approx 0.57\,.
\end{align}
This analysis should be understood as a demonstration that our framework can be implemented in GUTs, 
and as a reminder that left-right symmetric models with $D$-parity breaking allow for different gauge 
couplings $g_L \neq g_R$.  The latter fact can be used to explain the $eejj$ and dijet excesses at the LHC 
discussed in the next section. The scenario discussed here is not the only option to achieve differing left and right gauge couplings at the TeV scale. Other possibilities are for example explored in~\cite{Arbelaez:2013nga, Pelaggi:2015kna}.

\section{LHC signatures}
\label{sec:lhc}

\subsection{Experimental picture}
One of the most sensitive processes to probe Left-Right symmetric models at the LHC is given by heavy right-handed $W$ boson 
and neutrino exchange leading to the signal $pp\to W_R \to N l_1^\pm \to l_1^\pm l_2^{\pm,\mp} +2$~jets \cite{Keung:1983uu, Ho:1990dt, 
Ferrari:2000sp, Gninenko:2006br, 0nubb-LRa, Das:2012ii, AguilarSaavedra:2012fu, Deppisch:2014qpa}. Both the CMS and ATLAS collaborations 
have reported updated bounds on the mass of the right-handed charged gauge boson in the LRSM from their analyses of events 
at the center of mass energy $\sqrt{s} = 8$~TeV with an integrated luminosity of $\approx 20~{\rm fb}^{-1}$~\cite{cms_excess, 
Aad:2015xaa}. The two analyses treat events with two electron and two muons separately but they differ in that the CMS analysis 
does not differentiate between lepton charges, i.e. it includes both opposite sign and same sign leptons, whereas the ATLAS 
analysis only considers same sign lepton signatures as expected with a pure Majorana neutrino involved in the process. 
Observing no significant excess in either the $ee$, or $\mu\mu$ channel, both collaborations quote a lower limit at 95\% CL 
on the $W_R$ mass of the order $M_{W_R} > 3$~TeV, for $M_N \approx \frac{1}{2} M_{W_R}$.

The CMS data exhibit an excess in the $ee$ channel with a local significance of $2.8\sigma$ for a $W_R$ mass of $M_{W_R} 
\approx 2.1$~TeV. The excess is seen exclusively in the opposite sign channel $e^+ e^-$, with 13 potential signal events. 
Only one same sign electron event was observed~\cite{cms_excess}. Neither ATLAS nor CMS attempt to search for lepton flavor violating signatures; 
instead, the CMS analysis uses $e^\pm \mu^\mp$ events to determine the background from data. In the following, we attempt 
to interpret the CMS excess in our model. This seems especially interesting as there have been a number of additional 
(individually non-significant) excesses around the energy scale $M \approx 2$~TeV by both CMS and ATLAS.

When trying to interpret the excess in terms of resonant $W_R$ production and decay via an on-shell heavy neutrino, several 
issues have to be solved: 

(i) The $eejj$ excess is too small when compared to the predicted cross section in the minimal LRSM using $g_R = g_L$ 
by a factor of $\approx 3-4$. As first reported in \cite{Deppisch:2014qpa}, this issue is straightforward to understand 
in LRSMs without manifest left-right symmetry and $g_R \approx 0.6 g_L$ as is the case in our model. 

(ii) Neither ATLAS nor CMS see any excess in same sign lepton channels as expected in LRSMs with a pure Majorana 
neutrino where the number of same and opposite sign lepton events should be equal. This is easily reconciled in our model 
which incorporates Quasi-Dirac heavy neutrinos where the relative mass splitting $\approx m_\nu/M$ between the Majorana 
components is vanishingly small and only opposite sign lepton events are expected. 

(iii) There is no excess in the $\mu\mu jj$ channel. In fact CMS sees slightly fewer events than expected from the background 
for an invariant mass $M_{\mu\mu jj} \approx 1.5$~TeV. The phenomenologically most straightforward explanation for the lack 
of the $\mu\mu$ excess is the assumption that only one heavy neutrino is lighter than the $W_R$ and it dominantly couples 
to electrons. This would also mean that the right-handed currents are largely aligned with the charged lepton flavors 
and the heavy mixing matrix would be close to unity (neglecting the small non-unitarity). We will use this as a working 
assumption, although other solutions with non-trivial mixing (e.g. to the $\tau$ lepton) are possible 
\cite{Gluza:2015goa, Dobrescu:2015qna, Coloma:2015una} \footnote{By design, the CMS analysis does not report on the potentially possible 
flavor violating signature $e^\pm \mu^\mp jj$ as such events are used to infer the $t\bar t$ background. It would be interesting 
to see if dropping this assumption had any impact; if there were genuine $e^\pm \mu^\mp jj$ signal events, one might naively expect that the background is currently overestimated and correcting for this would result in a larger excess and might possibly reduce the under-fluctuation of the $\mu\mu jj$ 
event rate.}.

(iv) The CMS collaboration does not see a localized excess in the invariant mass distribution of the two jets and 
the sub-leading lepton $l_2$ as expected from the decay $N \to l_2 jj$. Although the leading and sub-leading lepton 
cannot be identified on an event-by-event basis, the leptons can be ordered by $p_T$, which should produce a characteristic 
excess distribution in $m_{l_2 jj}$. It is not clear how significant the absence of such a distribution is as its shape 
will depend on the mass of the heavy neutrino in the signal process and the low statistics might not allow to draw 
a conclusion. Another possible explanation is that the signal events are partially generated in another, kinematically 
different, signal process. An example would be that the neutrino decays in a left-handed current process via an on-shell 
SM $W$ boson as discussed below. Alternatively, a second neutrino with a different mass and non-trivially coupling to $\tau$ concurrently contributes 
to the signal \cite{Dobrescu:2015qna}. It is worthwhile to note that the CMS collaboration also does not see localized 
excesses in other distributions such as based on the invariant masses of other final state particles, e.g. $m^2_{qq}$ or 
$m^2_{l_1 l_2}$. 

(v) Last but not least, the reported excess of $2.8\sigma$ locally is currently not high enough to be statistically 
significant. Nevertheless, the combination with other LHC excesses around the resonant mass $M \approx 2$~TeV provides 
motivation to explore the observed excess despite its insufficient significance. Other analyses in the context of LRSMs 
were performed in~\cite{Heikinheimo:2014tba, Aguilar-Saavedra:2014ola, Dobrescu:2015qna, Brehmer:2015cia, Krauss:2015nba, Cheung:2015nha, Dhuria:2015cfa}. 
The excess has also been discussed in wider theoretical contexts~\cite{Bai:2014xba, Dobrescu:2014esa, Allanach:2014lca, 
Biswas:2014gga, others:2014, Cacciapaglia:2015nga, Chiang:2015lqa, Sanz:2015zha, Carmona:2015xaa, Anchordoqui:2015uea, Hisano:2015gna, Omura:2015nwa, Fichet:2015yia, Thamm:2015csa, Petersson:2015rza, Gao:2015irw, Dhuria:2015hta, Dhuria:2015swa}.

It is indeed intriguing to consider that all the various excesses around 2~TeV can be understood within left-right 
symmetric scenarios \cite{Deppisch:NuFermilabTalk, Dev:2015pga}. The different excesses can be summarized as follows:
\begin{enumerate}
\item \textbf{Diboson hadronic final states}: Both ATLAS and CMS have performed searches for a resonance hadronically 
decaying into a pair of the SM gauge bosons~\cite{Aad:2015owa, Khachatryan:2014hpa}. The jets emerging from the decays 
of the gauge bosons are nearly collinear and form a so called fat jet. The ATLAS search has a mild discrimination between 
fat jets emerging from the decays of $W$ and $Z$. Consequently, the results can be interpreted as decays of a resonance 
into $WZ$, $ZZ$ and $WW$ final states with overlapping events. ATLAS reports a local excess of $3.4 \sigma$, $2.9\sigma$ 
and $2.6 \sigma$ in these channels, respectively, in the region $\approx 1.9-2.1$~TeV, while CMS observes a $1.4\sigma$ 
excess at $\approx 1.9$~TeV with no discrimination between $W$ and $Z$ tagged jets. It should be noted that there are subtleties involved regarding the jet substructure in the experimental analysis, which could be improved in Run-2 as suggested in e.g.~\cite{Goncalves:2015yua}.

\item \textbf{Diboson semileptonic final states}: A CMS search~\cite{Khachatryan:2014gha} for a resonance decaying to SM gauge 
bosons with leptonically tagged $Z$ sees an excess of $1.5\sigma$ around 1.8~TeV.

\item  \textbf{Gauge boson - Higgs final state}: A CMS search~\cite{CMS-PAS-EXO-14-010} for a resonance decaying to SM $W$ 
and Higgs $H$, where $W$ decays leptonically and a highly boosted $H$ decays to pair of $b$ jets, sees an excess of $2.2\sigma$ 
around 1.8-1.9~TeV.

\item \textbf{Dijet final state}: Both ATLAS and CMS observe an excess in the dijet distribution of the decay of a resonance 
to two jets around 1.8~TeV with a significance of $1\sigma$ and $2.2\sigma$, respectively~\cite{Aad:2014aqa, Khachatryan:2015sja}.
\end{enumerate}
Although the above searches see an excess in the similar mass bins and hence generate a lot of interest, many other searches 
which are also sensitive to the decays of a $W_R$ do not see any excess:
\begin{enumerate}
\item\textbf{Diboson semileptonic final states}: ATLAS performed a search for a resonant diboson decay with subsequent 
leptonic decays of $W$ and hadronic decays of $Z$~\cite{Aad:2015ufa}. The search sees no excess of events around 2~TeV. 
ATLAS also performed a search in a diboson final state with leptonically decaying $Z$ and hadronically decaying 
$W$~\cite{Aad:2014xka} with no excess events seen. We do not attempt to explain the non-observation in this channel in our work. A possible way to reconcile results of this search with other excesses can be found in for example~\cite{Aguilar-Saavedra:2015rna}. 

\item\textbf{Gauge boson - Higgs final state}: Both ATLAS and CMS performed searches for a resonance decaying to a SM gauge 
boson and a Higgs. While ATLAS searches for leptonic decays of the gauge boson and the Higgs decaying to 
$b\bar{b}$~\cite{Aad:2015yza}, CMS searches for hadronic decays of the gauge boson and the Higgs decaying to 
$\tau\tau$ and $WW^{*}$ \cite{ Khachatryan:2015ywa, Khachatryan:2015bma}. No excess events were seen in either search.

\item\textbf{Third generation quarks final states}: Any heavy charged particle producing an excess in a dijet final 
state should also result in signal with third generation SM quarks. Both ATLAS and CMS performed searches for a resonance 
decaying to top and bottom final states, where the top decays either hadronically~\cite{Aad:2014xra} or 
semileptonically~\cite{Aad:2014xea, Chatrchyan:2014koa}. Neither of these searches report any excess of events. 
\end{enumerate}

\subsection{Theoretical prediction}
Understanding the diboson and dijet results in a common LRSM framework has been attempted the recent
work~\cite{Brehmer:2015cia}. In order to combine them with 
an interpretation of the CMS $eejj$ excess, we use the following values for the fitted cross sections derived 
from the LHC excesses~\cite{Brehmer:2015cia},
\begin{align}
\label{eq:fittedcs}
	\sigma(pp \to W_R \to WZ) &= 5.9^{+5.3}_{-3.5}\text{ fb}, \nonumber\\ 
	\sigma(pp \to W_R \to WH) &= 4.5^{+5.2}_{-4.0}\text{ fb}, \nonumber\\ 
	\sigma(pp \to W_R \to jj) &= 91^{+53}_{-45}\text{ fb}, \nonumber\\ 
	\sigma(pp \to W_R \to tb) &= 0^{+39}_{-0}\text{ fb}.
\end{align}
The numbers for the individual channels were derived by summing over the bins around 1.8-2.1~TeV 
and performing a cut and count analysis on this enlarged signal region. The input for producing 
these numbers are the number of observed events, expected backgrounds, efficiencies and systematic 
uncertainties, as published by ATLAS and CMS.

In addition to the analysis in \cite{Brehmer:2015cia}, we here also include the CMS $eejj$ excess in the analysis. We estimate the cross section $\sigma(pp \to W_R \to N e \to eejj)$ by taking into 
account the experimental efficiency $\epsilon = 0.754$ for $W_R\approx 1.9$~TeV and $N_R\approx 1.6$~TeV, 
cf. table A18 in \cite{cms_excess}. The observed number of events is 14 with an expected background of 
4 events. We assume all 10 signal events are within the resonance peak and hence derive a signal cross 
section of 0.66~fb. Clearly, the assumption that all signal events belong to the peak is an approximation. 
We also do not account for any systematic uncertainties or any other factors. However, in order to compensate 
for this, we assume a rather large error of 0.4~fb on the signal cross section,
\begin{align}
\label{eq:fittedeejj}
	\sigma(pp \to W_R \to N e \to eejj) &= 0.66^{+0.4}_{-0.4}\text{ fb}.
\end{align}
%

A detailed simulation of various processes is beyond the scope of this paper, but we attempt to demonstrate 
that the diboson and dijet excesses can be understood together with the CMS $eejj$ excess within the context 
of LRSMs. The total cross section for $W_R^+$ and $W_R^-$ can be expressed as \cite{Leike:1998wr}
\begin{align}
	\label{eq:cs_lhc_wr}
	\sigma(pp \to W_R) &=
	\frac{\pi}{12} \frac{g_R^2}{s}
	\left[
	    f_{u \bar d}\left(\frac{M_{W_R}}{\sqrt{s}}\right) 
	  + f_{d \bar u}\left(\frac{M_{W_R}}{\sqrt{s}}\right)
	\right].
\end{align}
The function $f_{q_1 q_2} = A_{q_1 q_2} \exp(-B_{q_1 q_2} M_{W_R}/\sqrt{s})$ approximates the PDF folded 
cross section of a resonant production with the fitting parameters in this case given by $A_{u\bar d} = 2750, 
B_{u\bar d} = 37$ and $A_{d\bar u} = 1065, B_{d\bar u} = 36$ \cite{Deppisch:2013jxa}. For $M_{W_R} = 1.9$~TeV 
and $\sqrt{s} = 8$~TeV, this yields the cross section
\begin{align}
	\sigma(pp \to W_R) = 390\text{ fb} \cdot \left(\frac{g_R}{g_L}\right)^2.
\end{align}
Right-handed $W_R$ bosons decay as $W_R \to q\bar q$, $eN$, $WZ$ and $WH$, the latter two decays through 
the suppressed $W$ boson mixing angle $\propto \sin^2\theta_{LR}^W$. We assume that there is only one 
heavy neutrino $N$ lighter than $W_R$ and there are no other particles below $M_{W_R}$. Neglecting SM particle 
masses kinematically, the respective partial decay widths as enumerated in~\cite{Brehmer:2015cia} are given as
\begin{align}
\label{eq:WPartialWidths}
	\Gamma(W_R^+ \to \sum_i q_i \bar{q}_i) &= 
		\frac{3g_R^2}{16\pi} M_{W_R}, \nonumber \\
	\Gamma(W_R^+ \to  e^+N) & = 
		\frac{g_R^2\,M_{W_R}}{48\pi}\left(1-\frac{M_N^2}{M_{W_R}^2}\right)^2 \!\!
		\left(1+\frac{1}{2}\frac{M_N^2}{M_{W_R}^2}\right),  \nonumber \\
	\Gamma(W_R^+ \to W^+ Z) &= 
		\frac{g_L^2}{192\pi}\sin^2\theta_{LR}^W\frac{M_{W_R}^5}{M_W^4},  \nonumber \\
	\Gamma(W_R^+ \to W^+ H) &= 
		\frac{g_L^2}{192\pi}\sin^2\theta_{LR}^W\frac{M_{W_R}^5}{M_W^4}.
\end{align}
The heavy neutrino decays for the $eejj$ signature through the three-body decay $N\to ejj$ 
via an off-shell $W_R$. Allowing for the potentially sizable $W$ boson ($\sin\theta_{LR}^W$) 
and LR neutrino ($\sin\theta_{LR}^N \equiv \Theta_{e1}$, i.e. the coupling of the only accessible 
neutrino to the electron due to LR mixing) mixing, it may also decay as $N \to eW$, $\nu_e Z$ and 
$\nu_e H$ \cite{Dev:2013oxa, Deppisch:2015qwa}. We assume that the heavy neutrinos are largely 
aligned with the charged lepton, i.e. the right-handed equivalent of the PMNS mixing matrix 
is close to unity and $V_{Ne} \approx 1$. In our calculations we treat the accessible heavy 
neutrino as a Dirac particle. Again neglecting SM particle masses kinematically, the respective 
partial decay widths as discussed in~\cite{delAguila:2008cj, Gluza:2015goa} are
\begin{align}
\label{eq:NPartialWidths}
	\Gamma(N \to e^- \sum_i q_i \bar{q}_i) &= 
		\frac{9g_R^4}{2048\pi^3}\frac{M_N^5}{M_{W_R}^4}, \nonumber\\
	\Gamma(N \to e^-W^+) &= 
		\frac{g_L^2 \sin^2\theta_{LR}^N + g_R^2 \sin^2\theta_{LR}^W}{64\pi} 
		\frac{M_N^3}{M_{W}^2}, \nonumber\\ 
	\Gamma(N \to \nu_e Z) &= 
		\frac{g_L^2\sin^2\theta_{LR}^N}{64\pi\,\cos^2\theta_W}\frac{M_N^3}{M_{Z}^2}, \nonumber\\ 
	\Gamma(N \to ZH) &= 
		\frac{g_L^2\sin^2\theta_{LR}^N}{64\pi}\frac{M_N^3}{M_{H}^2}. 
\end{align}
The above decay widths of $W_R$ and $N$ include all relevant couplings, especially the $W-W_R$ and $\nu-N$ mixing, consistently, which has not been done in previous analyses of the LHC excesses. Note, although we have given the formulas in the massless SM limit, we use the complete mass dependence in our calculation. It is also important to point out that we assume that the additional scalars in our 
model are heavier than $W_R$ and $N_R$, and hence their decays are not present. In case such mass 
hierarchy is not possible, the decays of $W_R,N_R$ to the heavy scalars should also be taken into 
account. 

\begin{figure}[t]
\centering
\includegraphics[width=0.9\columnwidth]{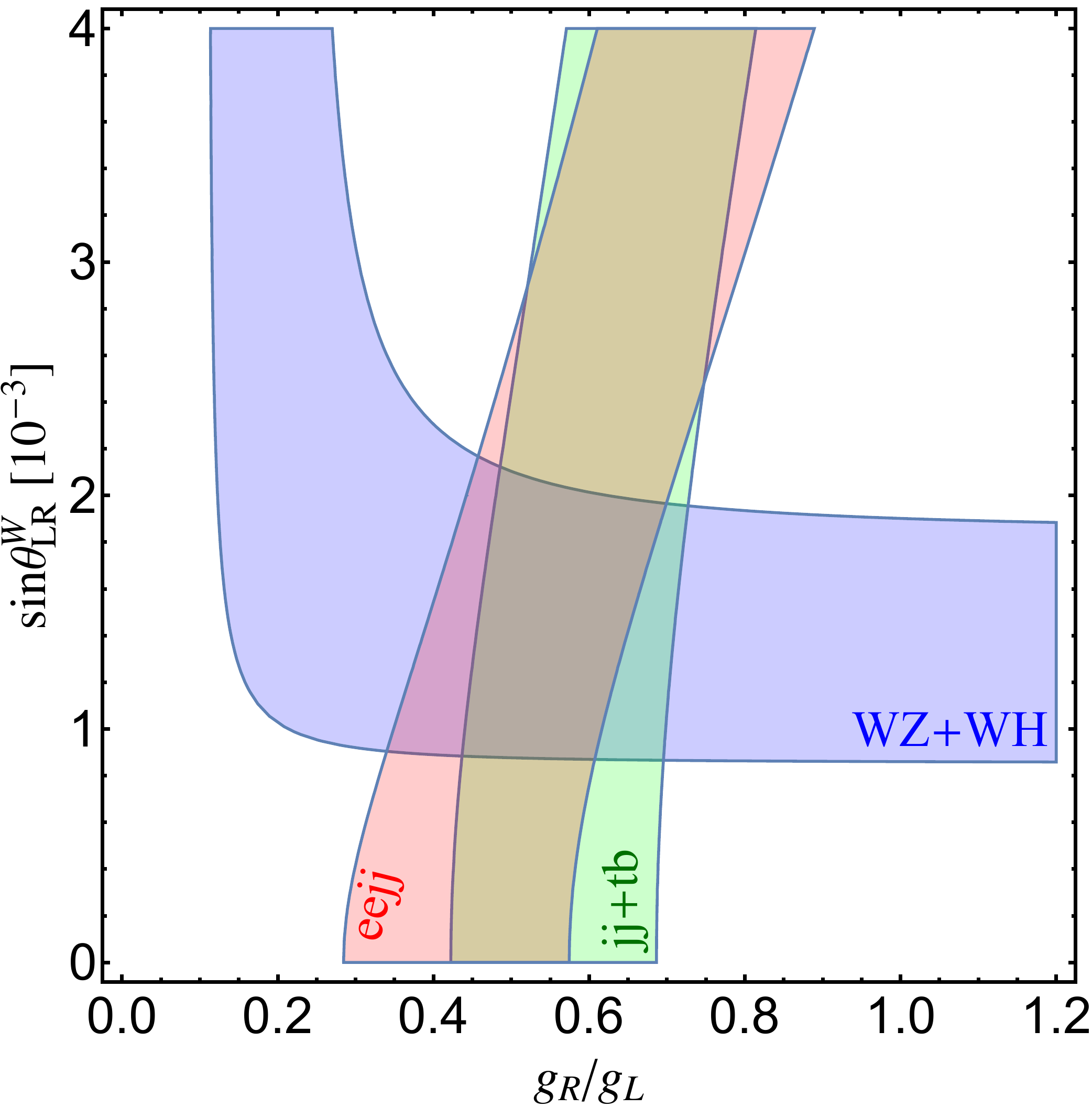}
\caption{Fitting the diboson $WZ, WH$ (blue band), dijet $jj+tb$ (green band) and $eejj$ (red band) excesses 
in the ($g_R/g_L$-$\sin\theta_{LR}^W$) parameter plane. The other parameters are chosen as $M_{W_R} = 1.9$~TeV, 
$M_N = 1.6$~TeV and $\sin\theta_{LR}^N = 0$.}
\label{fig:excessesLHC}
\end{figure}
Combining the $W_R$ production cross section and the decay widths, we can compare the experimentally suggested 
cross sections in Eqs.~\eqref{eq:fittedcs} and \eqref{eq:fittedeejj} with the theoretical predictions expressed 
in terms of the model parameters $g_R$, $\sin\theta_{LR}^W$, $\sin\theta_{LR}^N$ and $M_N$. We assume that the $W_R$ 
mass is given by $M_{W_R} = 1.9$~TeV. Figure~\ref{fig:excessesLHC} shows the compatibility of the theoretically 
predicted cross sections with the experimentally suggested ranges in the $g_R / g_L$ - $\sin\theta_{LR}^W$ parameter 
plane. The remaining parameters are chosen as $M_N = 1.6$~TeV and $\sin\theta_N = 0$. As can be seen and as has been 
noted in \cite{Brehmer:2015cia}, the dijet excess (assuming compatibility with the non-observation of $tb$) points 
to a right-handed gauge coupling with $g_R / g_L \approx 0.6$ whereas the diboson excesses suggest a sizeable $W$ boson 
mixing of $\approx 1.5\times 10^{-3}$. It is now very interesting to see from our analysis that the $eejj$ excess can be explained for the same values together with the diboson and dijet excesses.

\begin{figure}[t]
\centering
\includegraphics[width=0.9\columnwidth]{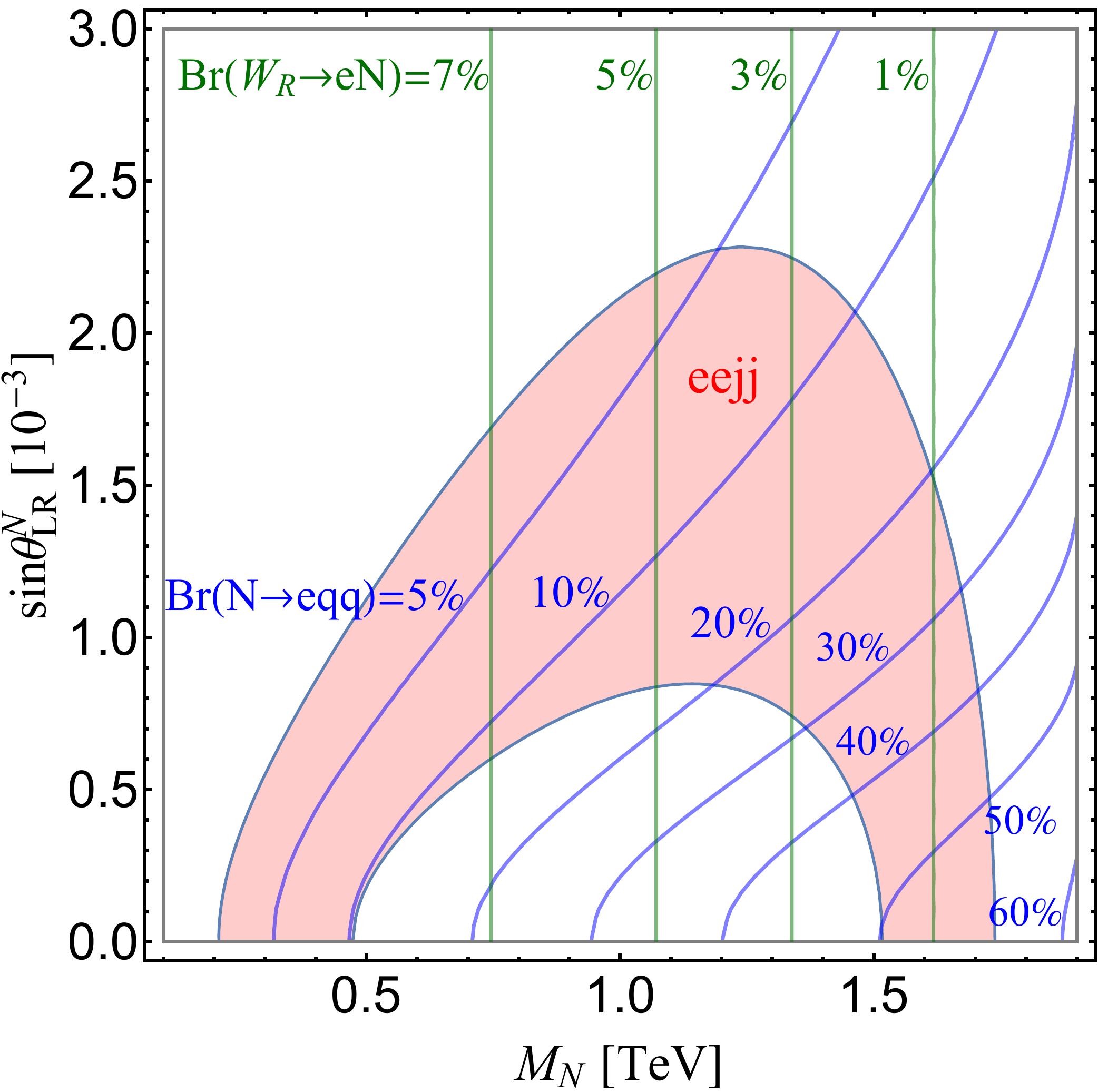}
\caption{Fitting the $eejj$ (red band) excess in the ($M_N$-$\sin\theta_{LR}^N$) parameter plane.  
The other parameters are chosen as $M_{W_R} = 1.9$~TeV, $g_R / g_L = 0.57$ and $\sin\theta_{LR}^W = 1.5\times 10^{-3}$. 
The vertical green lines denote contours of constant $\text{Br}(W_R \to e N_R)$ and the diagonal blue lines of constant 
$\text{Br}(N_R \to eqq)$ as denoted.}
\label{fig:excessesmNthetaN}
\end{figure}
The diboson and dijet cross sections only depend weakly on the neutrino mass and the LR neutrino mixing. On the other 
hand, the $eejj$ cross section delicately depends on these parameters through the branching ratios $\text{Br}(W_R \to e N)$ 
and $\text{Br}(N \to eqq)$. The latter is a three-body decay and thereby suppressed as $(M_N/M_{W_R})^4$. For a small 
neutrino mass, this decay has to compete with the two-body decays via left-handed current due to the small but sizable 
LR $W$ boson mixing and the potentially sizable LR neutrino mixing, cf. Eq.~\eqref{eq:NPartialWidths}. The compatibility 
of the predicted $eejj$ cross section with the observed excess is shown in Fig.~\ref{fig:excessesmNthetaN} as a red band 
in the ($M_N$-$\sin\theta_{LR}^N$) parameter plane. The other parameters are fixed to lie in the `best fit' intersection 
of Fig.~\ref{fig:excessesLHC}. The other excesses are compatible for this choice over the whole ($M_N$-$\sin\theta_{LR}^N$) 
parameter plane within uncertainties. Also shown are iso-contours for the branching ratios $\text{Br}(W_R \to e N)$ and $\text{Br}(N \to eqq)$, the product of which results in the shape of the red region. For negligible LR neutrino mixing $\sin\theta_{LR}^N \approx 0$, there are two solutions: a large neutrino mass $M_N \approx 1.6$~TeV (as used explicitly in Fig.~\ref{fig:excessesLHC}) and a small neutrino mass $M_N \approx 0.3$~TeV. In between these values, the excess could be explained through the inclusion of large LR neutrino mixing up to $\sin\theta_{LR}^N \approx 2\times 10^{-3}$, of the same order as the suggested LR $W$ boson mixing. We thus demonstrate that by interpreting the excesses in such a simplified LRSM would have profound impact on the properties of the heavy neutrino.

It is interesting to note that for such a sizable LR mixing, the heavy neutrino can dominantly decay via two-body left-handed current processes. The left-most blue contour in Fig.~\ref{fig:excessesmNthetaN} denotes a constant branching ratio $\text{Br}(N \to eqq) = 5\%$, i.e. with the remaining 95\%, the neutrino decays via a left-handed current process. This opens up the possibility that the $eejj$ final state is not only produced through $W_R \to e N \to ee W_R^* \to eejj$ but also $W_R \to e N \to ee W \to eejj$ via a SM $W$. While the kinematics and topology, and thereby the selection efficiency in the respective search, are very much different for the latter process, it may be able to `pollute' the $eejj$ signal. Although a very tentative conjecture, this may explain the absence of a localized excess in the $m^2_{e_2 qq}$ distribution of the $eejj$ search. This situation is especially expected to arise for lighter heavy neutrino masses, as this scenario favors the neutrino two-body decay and the 
resulting 
SM $W$ would not be so strongly boosted, increasing the chance of reconstructing the jets.

\section{Conclusions}
The excesses around the resonant energy of 2~TeV found in LHC searches are arguably the most promising hint for new physics 
to emerge at the LHC so far. While individually not very significant due to the large number of LHC searches, their coincidence 
at around 2~TeV provides motivation to interpret them. In this work we have attempted to do so in a left-right symmetric model 
where the excesses are produced through the decays of a right-handed $W_R$ boson with mass $\approx 2$~TeV.

The right-handed charged $W_R$ boson naturally arises in a left-right symmetric model with 
Higgs doublets and spontaneous $D$-parity breaking. Unlike manifest left-right symmetric models where the gauge couplings 
satisfy $g_R = g_L$, in the present case we have $g_R / g_L \approx 0.6$ around the TeV scale. This value is predicted due 
to $D$-parity breaking in an $SO(10)$ GUT embedding with Pati-Salam symmetry $SU(2)_L\times SU(2)_R \times SU(4)_C$ as its 
highest subgroup. The sub-eV neutrino masses are explained via a linear seesaw, where the suppression of the neutrino masses 
come from the high $D$-parity breaking scale. The heavy neutrinos are pseudo-Dirac to a high degree, which means that no signs 
of lepton number violation would be expected at the LHC. Indeed, 13 of the 14 events in the $pp\to eejj$ signal observed by 
CMS are opposite-sign electrons~\cite{cms_excess}. 

Due to the extra suppression of $g_R$ in our model, the $W_R$ production decreases at the LHC. This helps to understand the 
$eejj$ and $jj$ excesses as shown in Fig.~\ref{fig:excessesLHC}. Interpreting the diboson excesses as decays $W_R \to WZ, 
WH$ fixes the mixing of the $W$ bosons to $\sin\theta_{LR}^W \approx 10^{-3}$, a value largely compatible with the model 
prediction $\sin\theta_{LR}^W \approx (2 g_R/g_L) (M_W/M_{W_R})^2$. It is worthwhile to note that the indirect low energy 
bounds are also ameliorated in our model due to the smaller right-handed gauge coupling. For example, the strongest indirect 
bound on $M_{W_R}$ due to the $K_L - K_S$ mass difference is roughly given by $(g_R/g_L)^2 ((2.5\text{ TeV})/M_{W_R})^2 
\lesssim 1$~\cite{KLKS}. In our model, the limit weakens to $M_{W_R} \gtrsim 1.5$~TeV, compatible with the potential signal 
at $M_{W_R} \approx 2$~TeV as compared to the limit $M_{W_R} \gtrsim 2.5$~TeV in manifest LR symmetry.

The main information on the heavy neutrino sector comes from the interpretation of the $eejj$ excess through the decay $W_R \to e N$ 
where the heavy neutrino $N$ subsequently decays via an off-shell $W_R$ as $N \to ejj$. We work in a simplified scenario with 
a single heavy neutrino lighter than $W_R$ that has a potentially large mixing $\sin\theta_{LR}^N \lesssim 10^{-3}$ with 
the light neutrinos, cf. Fig.~\ref{fig:excessesmNthetaN}. We would like to highlight that such a large left-right mixing 
would induce large contributions to neutrinoless double beta decay via the so called $\lambda$ and $\eta$ diagrams. 
In fact, Eq.~\eqref{eq:halflives} show that the limits from neutrinoless double beta decay are of the same order or better 
for $M_{W_R} = 2$~TeV. Especially the $\eta$ contribution results in a stringent constraint on the mixing between the light and heavy neutrinos for $M_{W_R} \approx 2$~TeV and $\sin\theta_{LR}^W \approx 10^{-3}$. Taken at face value, the corresponding limit $\langle \theta \rangle^{0\nu} \lesssim 10^{-5}$ would restrict the parameter space to the bottom of Fig.~\ref{fig:excessesmNthetaN} with either $M_N \approx 0.3$~TeV or $M_N \approx 1.6$~TeV. In comparing LHC and neutrinoless double beta decay in this way we assume that the effective mixing parameters 
are of the same order, $\sin\theta_{LR}^N \sim \langle \theta \rangle^{0\nu}$, which could be violated for non-trivial flavor mixing structures. Large $W$ and neutrino mixing have 
the effect that at the LHC the purely right-handed current three-body decay of $N$ is competing with left-handed current 
decays. This can have important implications when interpreting the $eejj$ excess. 

We have performed the analysis of the recent excesses, including the CMS $eejj$ excess in an effective LRSM framework that incorporates a single heavy neutrino $N$ lighter than $W_R$ and that consistently allows for non-universal gauge couplings $g_R \neq g_L$ and potentially sizable $W-W_R$ and $\nu-N$ mixing. Our results, specifically Figs.~\ref{fig:excessesLHC} and \ref{fig:excessesmNthetaN} apply to other models in such a context.

If the excesses at around 2~TeV were to be confirmed by future data to originate from a $W_R$ in a left-right symmetric 
context, it would have profound implications. The reach of LHC Run-2 for LRSM models for 13 TeV has been explored in a number 
of papers~\cite{Brehmer:2015cia, Dobrescu:2015yba, Dobrescu:2015qna}. Due to a rapid increase in the production cross section 
of $W_R$, with a few fb$^{-1}$ of data, it should be possible to confirm the presence of a $W_R$ in the dijet channel. We have 
assumed that additional scalars in our model are heavier than $M_{W_R}$; shall this not be the case, the decays of $W_R$ via 
additional scalars can also be explored at 13~TeV. The production of a heavy $Z_R$ also present in the model will be difficult because the ratio of gauge couplings is close the theoretically allowed limit and the mass of $Z_R$ increases 
rapidly in this regime. Even disregarding the wider impact on Beyond-the-Standard Model physics and the path of gauge unification, 
it would directly affect neutrino physics and imply that $B-L$ violation and neutrino mass generation occurs at the TeV scale 
or below. Moreover, it will strongly disfavor models of high scale leptogenesis~\cite{Deppisch:2013jxa}. In such a case, 
the explanation of the baryon asymmetry of the universe could be found closely above or even below the electroweak 
scale \cite{postsp, nnbarrev, nnbarrev1}. Moreover, the $W_R$ predicted around 2 TeV can lead to an interesting dark matter 
phenomenology as explored recently \cite{julian_LR_DM}.

\section*{Acknowledgements}
The work of SP is partially supported by the Department of Science and Technology, Govt.\ of India under the financial grant SB/S2/HEP-011/2013 and partly by the Max Planck Society in the project MANITOP. NS is partially supported by the Department of Science and Technology, Govt. of India under the financial grant SR/FTP/PS-209/2011. US is partially supported by the J.C. Bose National Fellowship grant from the Department of Science and Technology, India. The work of WR is supported by the Max Planck Society in the project MANITOP. SK is supported by the `New Frontiers' program of the Austrian Academy of Sciences. SK thanks J. Tattersall for useful discussions. We thank T. Bandyopadhyay and M. Krauss for pointing out two typos in our paper.

\end{document}